\documentclass[aps,prb,reprint,superscriptaddress,amssymb,amsmath]{revtex4-1}
\usepackage{graphicx}
\usepackage{dcolumn}
\usepackage{bm, color}
\usepackage[none]{hyphenat} 
\usepackage{amsmath}

\begin{document}

\title{Large anomalous Hall and Nernst effect from nodal line symmetry breaking in Fe$_2$Mn$X$ ($X$=P,As,Sb)}
\author{Jonathan Noky}
\affiliation{Max Planck Institute for Chemical Physics of Solids, D-01187 Dresden, Germany}
\author{Claudia Felser} 
\affiliation{Max Planck Institute for Chemical Physics of Solids, D-01187 Dresden, Germany}
\author{Yan Sun} 
\email{ysun@cpfs.mpg.de}
\affiliation{Max Planck Institute for Chemical Physics of Solids, D-01187 Dresden, Germany}

\date{\today}

\begin{abstract}
A large Berry curvature in the vicinity of the Fermi energy is required in order to obtain a large anomalous Hall and Nernst effect. This Berry curvature can be induced by Weyl points and gapped nodal lines. One of the possible mechanisms takes place in systems with a symmetry group where mirror planes lead to protected nodal line structures. When these mirror symmetries are broken, e.~g. via fixing a magnetization direction, a gap in the former nodal line can create Weyl points and a large Berry curvature along the gapped lines. In this work we study this effect in a minimal effective model and relate the results to the real regular Heusler compounds Fe$_2$Mn$X$ ($X$=P,As,Sb). These materials have three mirror planes in the non-magnetic case, leading to three nodal lines near the Fermi level. However, dependent on the orientation of the magnetization, some of the mirror planes are broken and the respective lines are gapped, creating large Berry curvature. Because the Fermi level is located in vicinity of the gapped lines, this leads to a large anomalous Hall and Nernst effect, which can be tuned to even higher values with a little bit of doping in the system.
\end{abstract}
\maketitle

\section{Introduction}

In recent years, the search for materials with large anomalous transport properties has raised increased interest, especially for topological (semi-)metals\cite{Manna2018}. These systems often show a strong anomalous Hall effect (AHE)~\cite{Pugh_1953,Lee_2004, XiaoDi_2006,WBauer2012,Nakatsuji2015,liu2017anomalous,Kiyohara_2016,PhysRevLett.118.136601,liang2018anomalous,Nagaosa_2010}, which is observed as a transverse voltage drop as a consequence of a longitudinal current induced only by the intrinsic spin-orbit coupling and magnetism. Additionally, very recently there is also interest in the anomalous Nernst effect (ANE)~\cite{Nerst_1887,WBauer2012,Lee_2004, XiaoDi_2006,guo2017large,PhysRevLett.119.056601,Ikhlas_2017}, which is analogous to the AHE when exchanging the longitudinal driving current with a temperature gradient.

The anomalous transport coefficients are directly linked to the Berry curvature (BC) of the electronic band structure near the Fermi level~\cite{Nagaosa_2010,Xiao2010, XiaoDi_2006}. This implies that potential material candidates have to have strong BC in their electronic structure. To create a large BC in a band structure, topological properties are neccessary. One way to achieve this is to look for a system with symmetry protected nodal lines and then break this symmetry. This removes the protected degeneracy and the line will gap out with a possibility of preserved crossings which are Weyl points. In this case the BC will flow from one Weyl point to the other via the now gapped line structure, inducing large BC at these lines. This was recently observed in Co$_2$MnGa~\cite{Guin2018,Sakai2018} where it leads to a large anomalous Hall and Nernst conductivity. 

The strength of the effect strongly depends on the energetic distance of the nodal line structure from the Fermi level. This makes it neccessary to either find systems where the Fermi level has a suitable value or take systems with interesting topological features and dope them to the right energy level. For the second approach, which grants more versatility, a promising class of materials to look at are the Heusler compounds where the Fermi level can easily be varied through the stochiometric substitution of single atoms~\cite{}. 
There are many studies showing a large ANE in both non-collinear~\cite{Chen_2014,Kubler_2014,Nakatsuji2015,Kiyohara_2016,Nayak2016,Ikhlas_2017,Li_2017} and collinear magnetic Heusler compounds~\cite{Shi_2018,Noky2018,Guin2018,Sakai2018}.

In this work we present a detailed analysis of the nodal line symmetry breaking mechanism described above for a minimal effective model and the real Heusler compounds Fe$_2$Mn$X$ ($X$=P,As,Sb). Here, the symmetry, which protects the nodal lines, is broken due to the orientation of the magnetization in these compounds. Additionally, the Fermi level is located at the nodal line structure, leading to a large anomalous Hall conductivity (AHC) and some of the highest reported values for the anomalous Nernst conductivity (ANC). Both of these effects can be enhanced to even higher values when doping the system to a suitable Fermi level.

\section{Methods}

We start the theoretical investigation with an \textit{ab initio} calculation of the electronic band structure via density-functional theory (DFT). For this we employ the \textsc{VASP} package~\cite{kresse1996} with a plane-wave basis set and pseudopotentials. The exchange-correlation potential is described in the generalized-gradient approximation (GGA)~\cite{perdew1996}. In the next step, we extract Wannier functions from the DFT band structure via the \textsc{Wannier90} package~\cite{Mostofi2008}. The initial projections are chosen as s-, p-, and d- orbitals of Fe and Mn and the s- and p-orbitals of $X$. From the Wannier functions we construct a Tight-Binding Hamiltonian, $H$, to calculate the BC, $\Omega$, in the system via the Kubo formula~\cite{Thouless_1982,Xiao2010,Nagaosa_2010}
\begin{equation}
  \Omega_{ij}^n=i\sum_{m\ne n} \frac{\langle n|\frac{\partial H}{\partial k_i}|m\rangle \langle m|\frac{\partial H}{\partial k_j}|n\rangle - (i \leftrightarrow j)}{(E_n-E_m)^2}.
\end{equation}
Here, $|n\rangle$ and $E_n$ are the eigenstates and eigenenergies of $H$, respectively.

From this we evaluate the AHC, $\sigma$, via
\begin{equation}
 \label{eq:ahc}
 \sigma_{ij}=\frac{e^2}{\hbar} \sum_n\int \frac{d^3k}{(2\pi)^3}\Omega_{ij}^nf_n
\end{equation}
and the ANC, $\alpha$, via the equation proposed by Xiao et al. \cite{Xiao2010, XiaoDi_2006}
\begin{align}
  \label{eq:anc}
  \alpha_{ij}=-\frac{1}{T} \frac{e}{\hbar} \sum_n \int \frac{d^3k}{(2\pi)^3} \Omega_{ij}^{n}[(E_{n}-E_F)f_{n}+\nonumber \\
  +k_BT\ln{(1+\exp{\frac{E_{n}-E_F}{-k_BT}})}].
\end{align}
Here, $T$ is the actual temperature, $f_n$ is the Fermi distribution, and $E_F$ is the Fermi level.

\section{Results}

\begin{figure}[htb]
\centering
\includegraphics[width=0.48\textwidth]{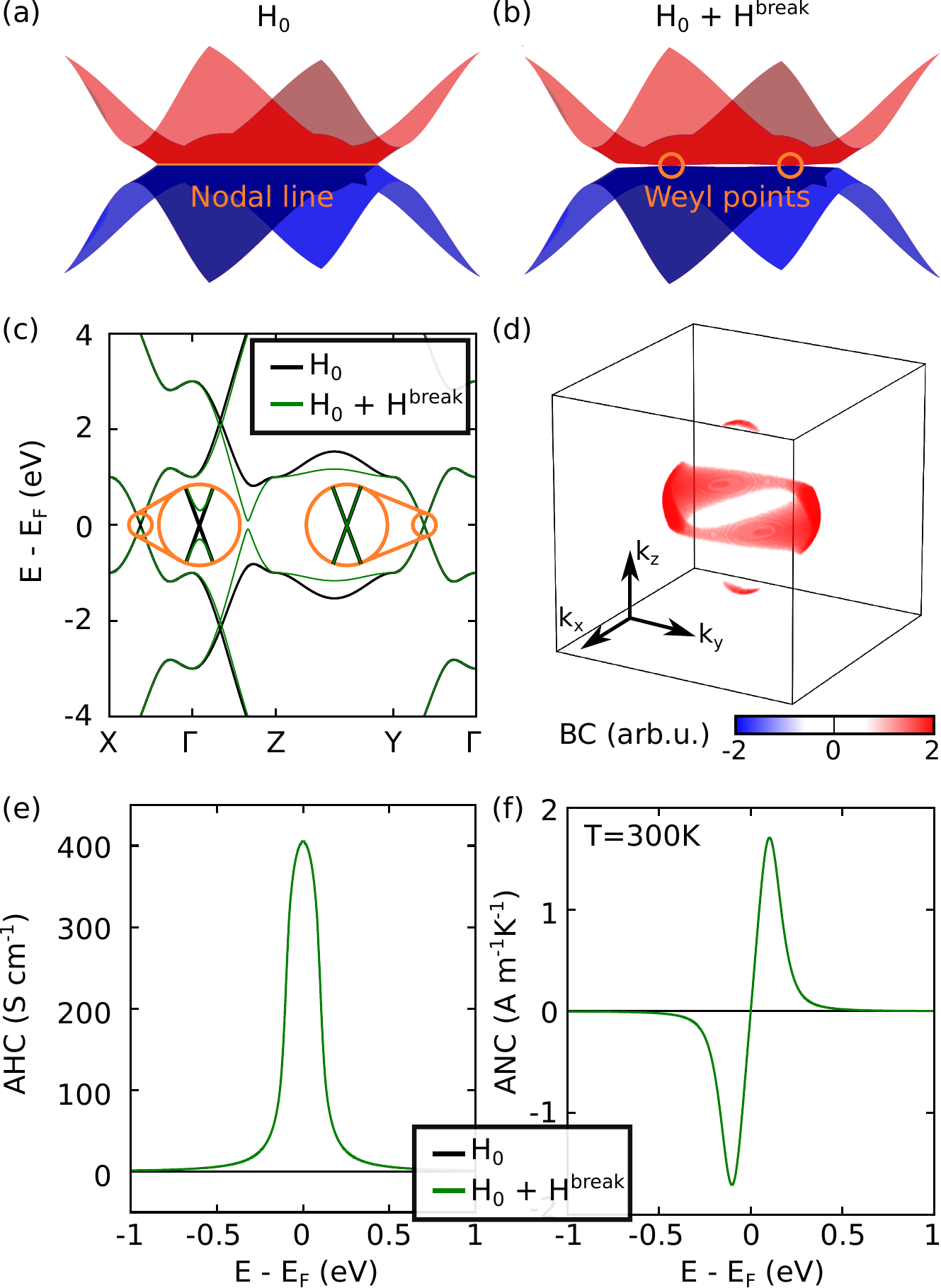}
   \caption{(a) Nodal line in the $k_x$-$k_y$ plane protected by the mirror symmetry in this plane. (b) Gapped line in the $k_x$-$k_y$ plane due to breaking the mirror symmetry with two Weyl points. (c) Band structure with and without the mirror symmetry. The insets show the splitting in the $\Gamma$-$X$ direction. (d) Berry curvature in the Brillouin zone after breaking the mirror symmetry. It is mostly located along the former nodal line. (e) Anomalous Hall conductivity. (f) Anomalous Nernst conductivity at 300 K.}
\label{fig:1}

\end{figure}

Firstly, we study the properties of a model system with a symmetry protected nodal line at the Fermi level. For this we employ the model proposed by Rauch et al.~\cite{PhysRevB.96.235103}
\begin{align}
 H_0=(m-6M&+2(\cos{k_x}+\cos{k_y}+\cos{k_z}))\tau_z\otimes\sigma_0+\nonumber\\ &+B\tau_z\otimes\sigma_z+c\sin{k_z}\tau_x\otimes\sigma_z\nonumber\\&+c\sin{k_x}\tau_x\otimes\sigma_x+c\sin{k_y}\tau_x\otimes\sigma_y.
\end{align}
This model system shows a nodal line protected by a mirror symmetry in the $k_x$-$k_y$ plane at $E_F$, when choosing the parameters as $m=M=c=1$ eV and $B=2$ eV~\cite{PhysRevB.96.235103}. The corresponding band structure is shown in Fig.~\ref{fig:1}(a) and (c). It is important to note that the nodal line induces only local BC, the total value obtained via an integration as described in equation (\ref{eq:ahc}) vanishes. As a result, this always leads to zero AHC and ANC for the closed nodal line system. Consequently, to get non-zero anomalous transport coefficients it is neccessary to gap out the nodal line. This is only possible when the mirror symmetry, which protects the nodal line, is broken. For this we add a small additional term $H^{break}=0.1c(\sin{k_x}\tau_x\otimes\sigma_x+\sin{k_z}\tau_x\otimes\sigma_y)$ to the Hamiltonian which can be for example spin-orbit coupling. As a consequence, the nodal line gaps out except for the two points with $k_y=0$ which are then Weyl points, as shown in Fig.~\ref{fig:1}(b). This gapping also induces strong BC into the band structure which can be seen in Fig.~\ref{fig:1}(d). It is obvious that the BC is generated along the former nodal line in the $k_z=0$ plane. Consequently, this BC leads to a non-zero AHE (Fig.~\ref{fig:1}(e)) and creates an ANE, although at the energy of the former nodal line this effect vanishes due to conserved particle-hole symmetry~\cite{Noky2018} (Fig.~\ref{fig:1}(f)). The model shows the importance of mirror symmetries for nodal line semimetals and reveals possible effects when these symmetries are broken~\cite{}. In the following we present real materials which show the same behaviour as observed in the model.

\begin{figure}[htb]
\centering
\includegraphics[width=0.48\textwidth]{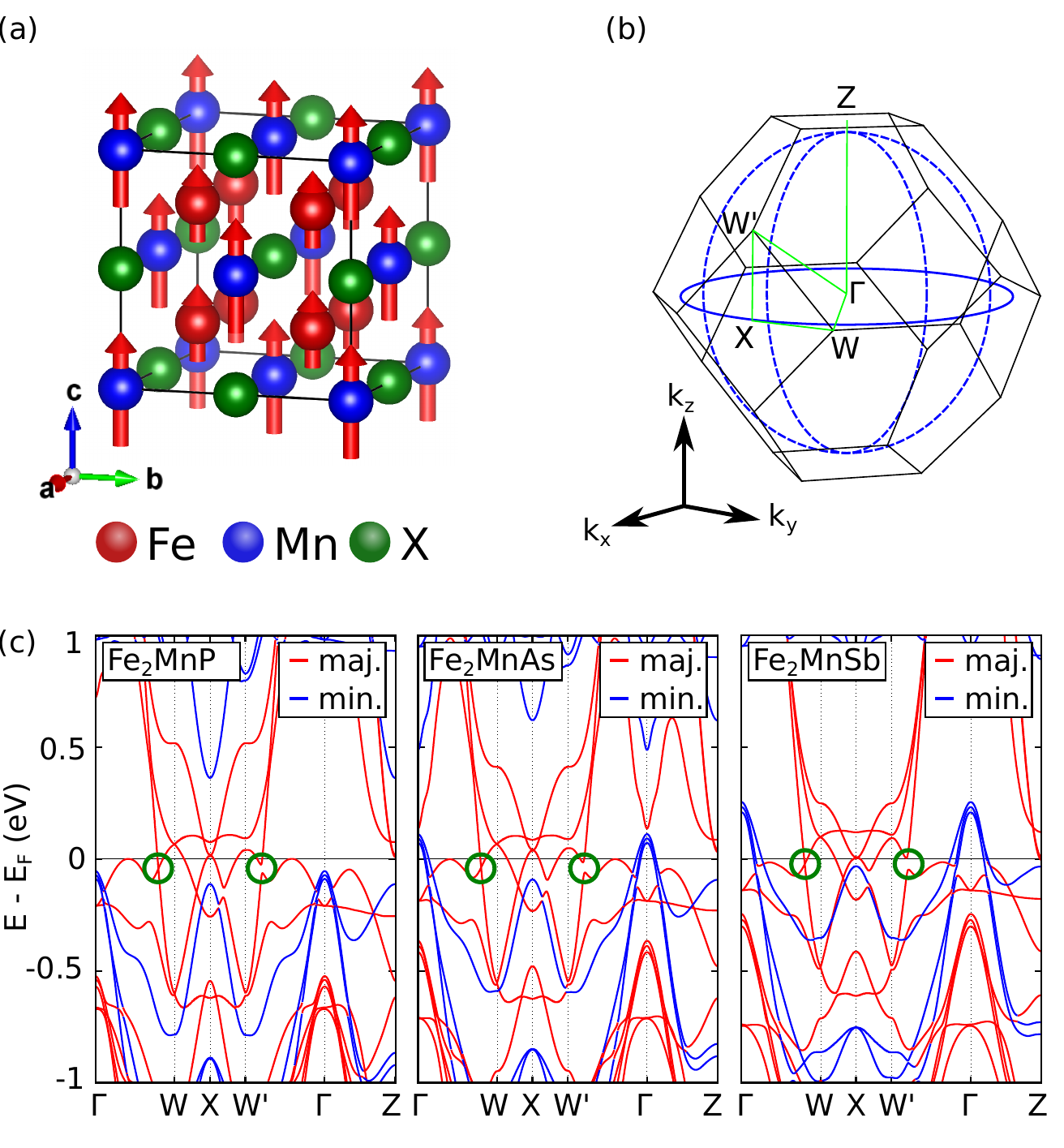}
   \caption{(a) Crystal structure of Fe$_2$Mn$X$ with magnetic moments. (b) Brillouin zone with path for the band structure (green lines) and nodal lines (blue lines). The dashed lines become gapped due to the chosen magnetization direction. (c) Band structures with spin-orbit coupling for Fe$_2$MnP, Fe$_2$MnAs, and Fe$_2$MnSb. The color of the bands mark a positive (red) or negative (blue) expectation value of $\sigma_z$. The green circles mark the positions of the nodal lines, where the first belongs to the still protected mirror in the $x$-$y$ plane and the second is gapped out.}
\label{fig:2}
\end{figure}

We investigate the regular Heusler compounds Fe$_2$Mn$X$ ($X$=P,As,Sb) with space group $Fm\overline{3}m$ (No. 225). In Fig.~\ref{fig:2}(a) the crystal and magnetic structure is shown. The compounds are ferromagnetic with a net magnetic moment of $\mu=4\mu_B$ per unit cell. We choose this to be aligned along the (001) direction with moments at the single atoms of $\mu_{Fe}=0.7\mu_B$, $\mu_{Mn}=2.6\mu_B$, and $\mu_X=0$. The crystal structure without magnetization possesses three mirror planes at $k_x=0$, $k_y=0$, and $k_z=0$. In each of these planes there is a mirror symmetry protected nodal line at the Fermi level (see Fig.~\ref{fig:2}(c)). When the magnetization is set along the (001) direction and spin-orbit coupling plays a role, the mirror symmetries in the $k_x=0$ and $k_y=0$ planes break. As a consequence, these lines gap out, analogous to the model described above. Furthermore, also here a large BC is induced into the band structure. This large BC strongly influences the AHC and ANC, as it can be seen from equations (\ref{eq:ahc}) and (\ref{eq:anc}). It is important to note, that a change of the magnetization orientation will change how the  nodal lines gap. This is, because the protective mirror symmetries are broken differently. However, even if the local BC in the system is changed, the total BC which results in the AHC and ANC stays the same for all magnetization directions in these systems~\cite{Guin2018}.

\begin{figure}[htb]
\centering
\includegraphics[width=0.48\textwidth]{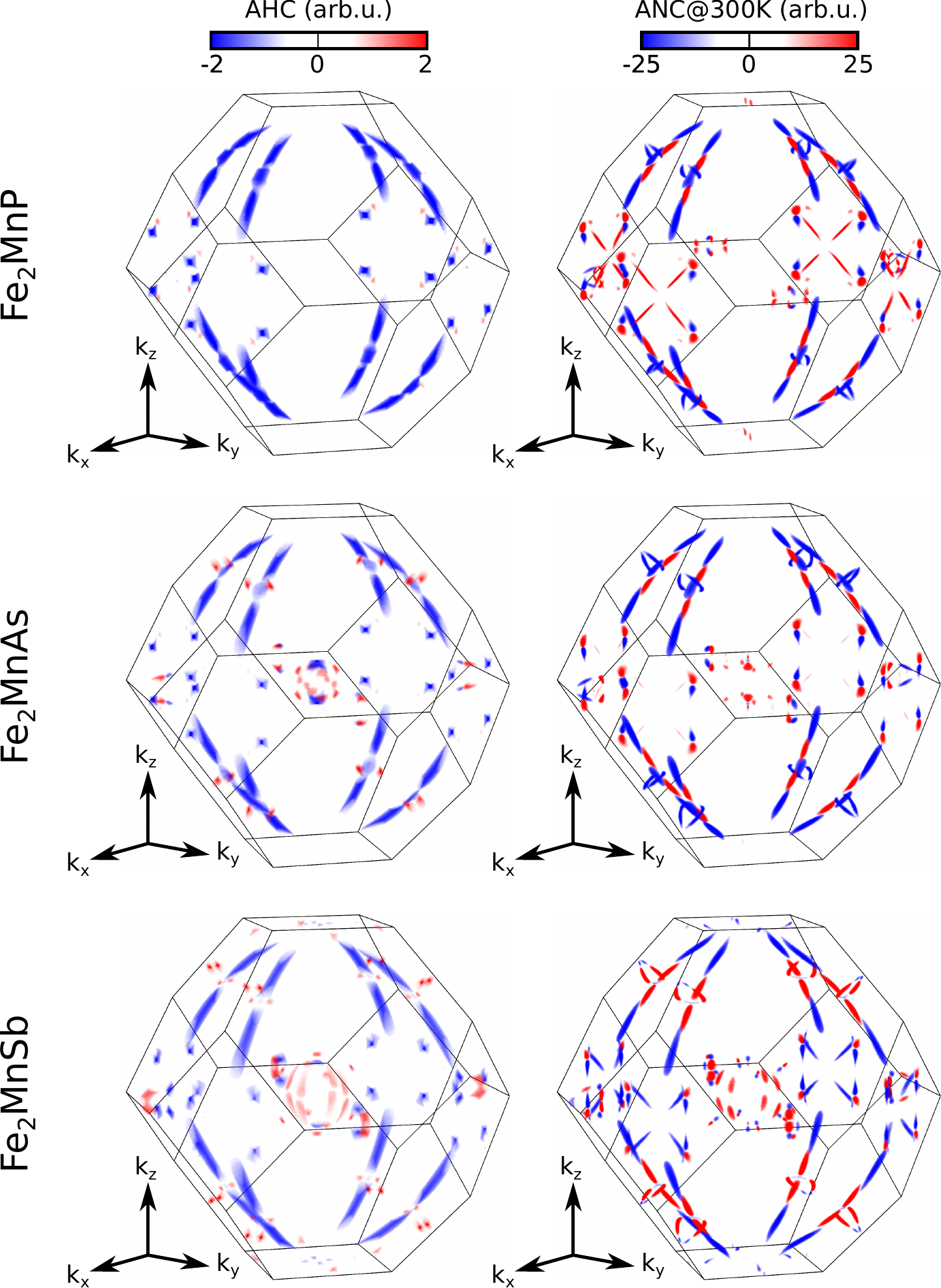}
   \caption{Contributions to the anomalous Hall and Nernst effect in the 3D Brillouin zone for Fe$_2$Mn$X$ ($X$=P,As,Sb). The largest contributions stem from the gapped nodal lines in the $k_x$-$k_z$ and $k_y$-$k_z$ plane. The $k_x$-$k_y$ plane does not contribute because in this plane the mirror symmetry is still valid, resulting in a closed nodal line.}
\label{fig:4}
\end{figure}

In Fig.~\ref{fig:4} the contributions to both AHC and ANC in all three compounds are shown  as a function of the $k$ space position. It can be seen that the main contributions stem from line-like shapes in the $k_x=0$ and the $k_y=0$ plane which coincide with the two broken mirror symmetry planes. Therefore, just as in the model, when the mirror symmetry which protects the nodal line is broken, the former nodal line gaps out and strongly enhances the BC at this gap.

\begin{figure}[htb]
\centering
\includegraphics[width=0.48\textwidth]{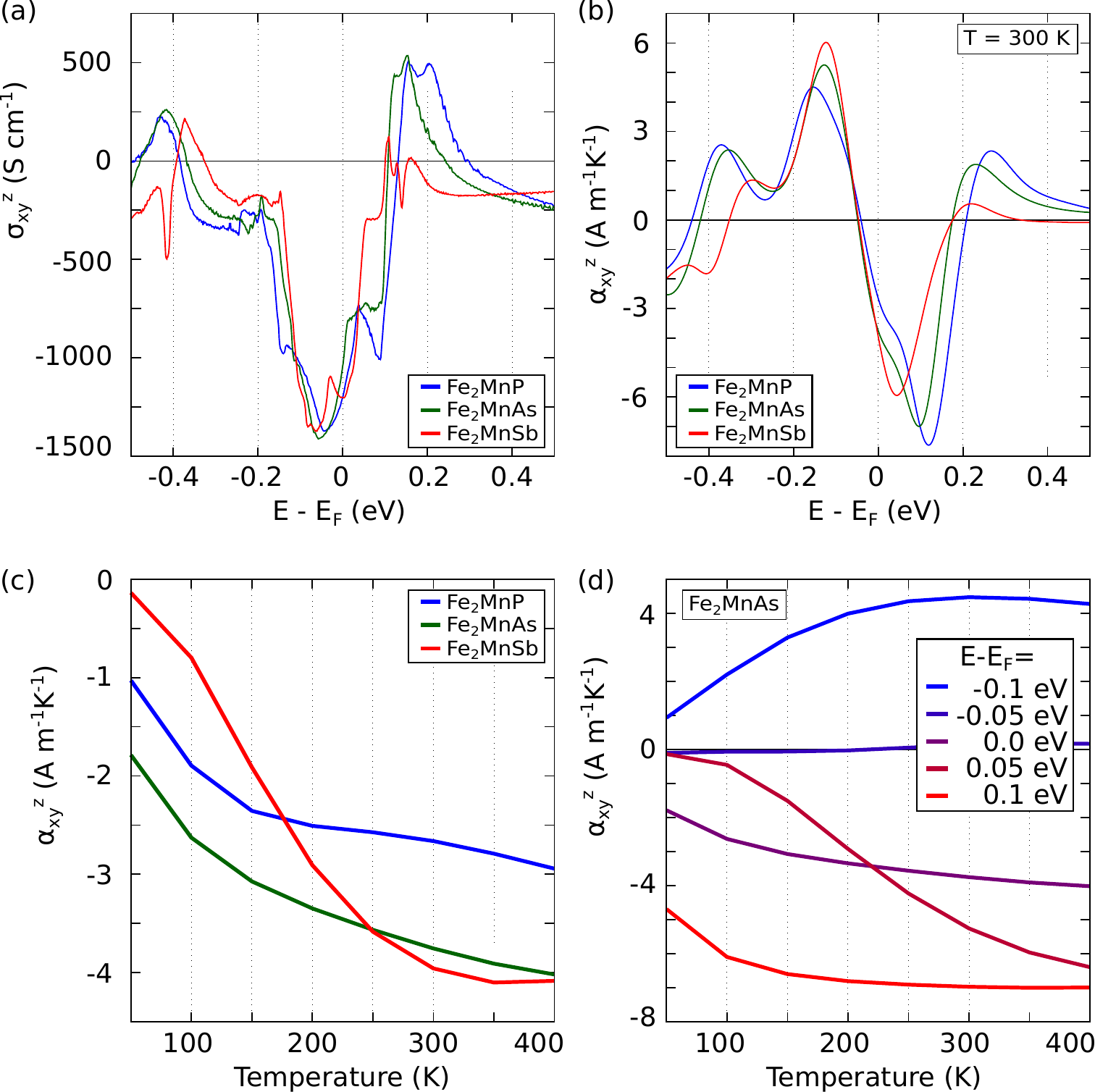}
   \caption{(a) Anomalous Hall effect of Fe$_2$Mn$X$. (b) Anomalous Nernst effect  of Fe$_2$Mn$X$ at $T=300$ K in dependence on energy level. (c) Anomalous Nernst effect in dependence on temperature. (d) Anomalous Nernst effect of Fe$_2$MnAs depending on temperature and energy level.}
\label{fig:3}
\end{figure}

This strong enhancement now also leads to large overall anomalous transport coefficients. In Fig.~\ref{fig:3}(a) the energy dependent AHC for Fe$_2$Mn$X$ ($X$=P,As,Sb) is shown. As is can already be expected from the similar band structures in Fig.~\ref{fig:2}(c) the AHC looks very similar for all three compounds. The value at the Fermi level is quite large with $\sigma_{xy}^z=-1200$ S cm$^{-1}$ for $X$=P,Sb and $\sigma_{xy}^z=-1050$ S cm$^{-1}$ for $X$=As. This is also near the maximum value for the AHC which is located slightly below the Fermi level. Remarkably, the ANC shown in Fig.~\ref{fig:3}(b) has a value of $\alpha_{xy}^z=-2.6(-3.8,-4)$ A m$^{-1}$ K$^{-1}$ for  Fe$_2$MnP(As,Sb), respectively, at the Fermi level and at T = 300 K. An analysis of the temperature dependence in Fig.~\ref{fig:3}(c) shows a slight increase of these values with increasing temperature. Due to the fact that Heusler compounds can be doped to different Fermi energies relatively easy~\cite{} it is also interesting to investigate the ANC for different positions of the Fermi level and their temperature dependence. As shown in Fig.~\ref{fig:3}(d) for Fe$_2$MnAs (the other compounds show a similar behaviour) it is possible to increase the ANC via doping by a factor of two at $E=E_F-0.1$ eV, which is also very stable in a large range of temperature. This doping level corresponds to 0.33 additional electrons per unit cell. It is important to note, that with a doping of $E=E_F-0.05$ eV (-0.18 electrons per unit cell) the ANC almost completely vanishes in the whole temperature range. This could lead to problems where interesting topological features are overlooked when randomly doped samples are investigated.

\section{Summary}

In summary, we have theoretically investigated the regular Heusler compounds Fe$_2$Mn$X$ ($X$=P,As,Sb). These materials show a nodal line structure, which gets gapped due to the symmetry breaking induced by the fixing of the magnetization direction. This breaking strongly enhances the Berry curvature in the system along the gapped lines and, consequently, leads to very large anomalous Hall and Nernst conductivities. With the help of a simple model we explained the underlying mechanisms and showed their signatures in the real materials. This work provides an intuitive understanding of the high anomalous transport coefficients in magnetic nodal line materials, especially in regular Heusler compounds, in terms of the breaking of protective symmetries.

We thank Jacob Gayles for the helpful discussions. This work was financially supported by the ERC Advanced Grant No. 291472 'Idea Heusler' and ERC Advanced Grant No. 742068 'TOPMAT'.

\end{document}